\documentclass[twocolumn,showpacs,preprintnumbers,amsmath,amssymb]{revtex4}

\usepackage{graphicx}
\usepackage{dcolumn}
\usepackage{bm}
\begin{document}
\title{ Electronic instability in bismuth far beyond the quantum limit}
\author{Beno\^{\i}t Fauqu\'e$^{1}$, Baptiste Vignolle$^{2}$, Cyril Proust$^{2}$,
Jean-Paul Issi$^{3}$ and Kamran Behnia$^{1}$}

\affiliation{(1)LPEM
(UPMC-CNRS), ESPCI, Paris, France\\ (2)Laboratoire National de
Champs Magn\'etiques Intenses (CNRS-INSA), Toulouse, France\\
(3) CERMIN, Universit\'e Catholique de Louvain, Louvain-la-Neuve, Belgium}

\date { September 10, 2009}

\begin{abstract}
We present a transport study of semi-metallic bismuth in presence of a magnetic field
applied along the trigonal axis extended to 55 T for electric conductivity
and to 45 T for thermoelectric response. The results uncover a new field scale at about 40 T
in addition to the  previously detected ones. Large anomalies in all transport
properties point to an intriguing electronic instability deep in the ultraquantum regime.
Unexpectedly, both the sheer magnitude of conductivity and its metallic temperature dependence are
\emph{enhanced} by this instability.

\end{abstract}


\maketitle

\begin{figure}
  \resizebox{!}{0.62\textwidth}{\includegraphics{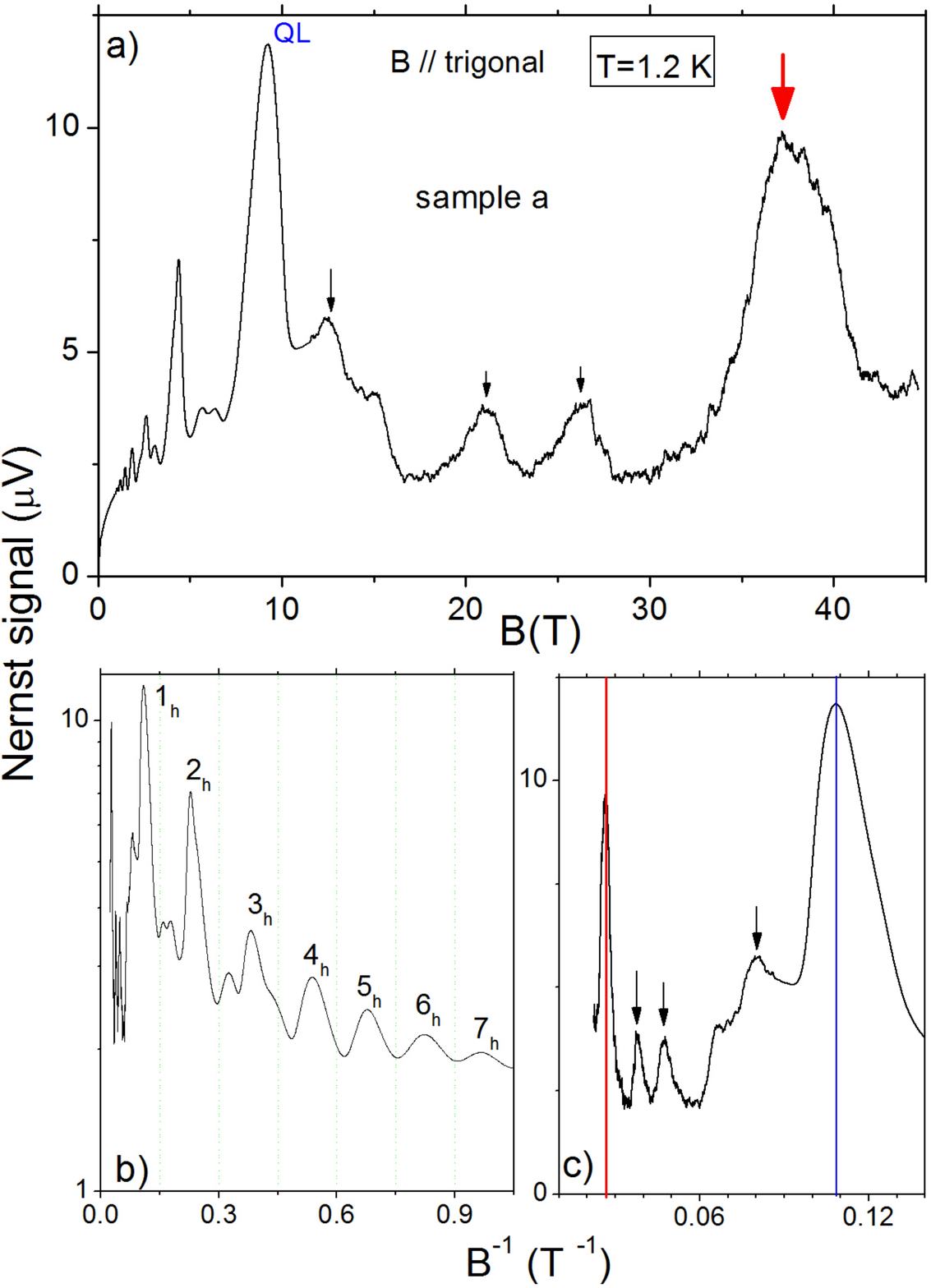}}
\caption{\label{Fig1} a) The Nernst signal as function of magnetic field. Above the quantum limit (QL), a large Nernst peak (large red arrow) is resolved in addition to the anomalies previously reported  (small black arrows). b) The same data  in a semi-log plot as a function
of B$^{-1 }$. Low-field quantum oscillations  are visible. The Nernst peaks are indexed as explained in the text. c) A zoom (in linear scale) on the high-field regime with vertical lines marking the two main peaks.}
\end{figure}

In presence of a reasonably large magnetic field, the electric resistivity of bismuth enhances by five orders of magnitude with no sign of saturation. This feature, discovered as early as 1928\cite{kapitza}, remains beyond the scope of the semiclassical transport theory, but has escaped wide theoretical attention\cite{abrikosov}. When the field is applied along the trigonal axis of the crystal and exceeds 9 T\cite{bompadre,behnia2}, no more crossing of the chemical potential by any known Landau level is expected. However, recent experimental studies of bismuth\cite{behnia1,li,fauque} have uncovered a number of enigmatic field scales beyond this quantum limit, which have attracted renewed theoretical attention to the energy spectrum of bismuth\cite{alicea,sharlai} and to the possible realization of the Fractional Quantum Hall Effect in  three dimensions\cite{burnell,levin}.

Two independent set of recent theoretical calculations\cite{alicea,sharlai} have both found that one-particle spectrum of bismuth for a field oriented close to the trigonal axis is remarkably complex due to the implications of the charge neutrality in a compensated system, the particular Fermi surface topology  and the relatively large Zeeman energy. According to these calculations, the field scales of the three electron ellipsoids are expected to present a very sharp angular dependence when the field is slightly tilted off the trigonal axis\cite{alicea,sharlai} as found by the torque magnetometry experiments\cite{li}. On the other hand, no other field scale above 11 T is expected in the one-particle picture, if the field is strictly oriented along the trigonal axis. In this context, the three high-field anomalies observed in the Nernst response above this field\cite{behnia1}, unexplained in the one-particle picture\cite{alicea,sharlai}, could be attributed to the Landau levels of the electron ellipsoids assuming a small misalignment of a few degrees\cite{sharlai}. Because of the absence of \emph{in situ} orientation of the crystal in the Nernst experiment\cite{behnia1}, this possibility could not be ruled out .

In this letter, we report on two sets of studies which extend the field range of the data up to 45 T (for the Nernst effect) and up to 55 T (for the resistivity tensor). The measurements uncover a new field scale in the vicinity of 40 T pointing to an unidentified electronic instability. The signatures of electronic reorganization at this field in experimental data, the drop in electric resistivity and the enhancement in the Nernst response, are almost as drastic as the crossing of the quantum limit at 9 T. Our angular-dependent resistivity studies establish that this field scale persists even when the field is strictly parallel to the trigonal axis and, thus cannot be attributed to the one-particle energy spectrum of electron ellipsoids.  Thus, it provides the most solid experimental evidence for a field-induced electronic instability caused by electronic interactions in bulk bismuth. Intriguingly, this electronic reorganization leads to a better conductivity and an enhanced metallicity, in sharp contrast to the expected signatures of a density-wave transition, widely believed to occur in graphite in a similar configuration.

 Bismuth single crystals of millimetric dimensions and Residual Resistivity Ratios ($\rho$(300K)/ $\rho$(4.2K)) ranging from 40 to 110  were used in this study. Nernst effect was measured by a miniature one-heater-two-thermometer set-up specially designed to work in the 45 T hybrid magnet of the  NHMFL-Tallahassee. Resistivity and Hall effect were measured in a pulsed magnet in LNCMI-Toulouse. For the angular-dependent resistivity measurements, we used a rotator which allowed to control with a precision of 2 degrees the relative orientation of the magnetic field and the crystal axes.

\begin{figure}
\resizebox{!}{0.62\textwidth}{\includegraphics{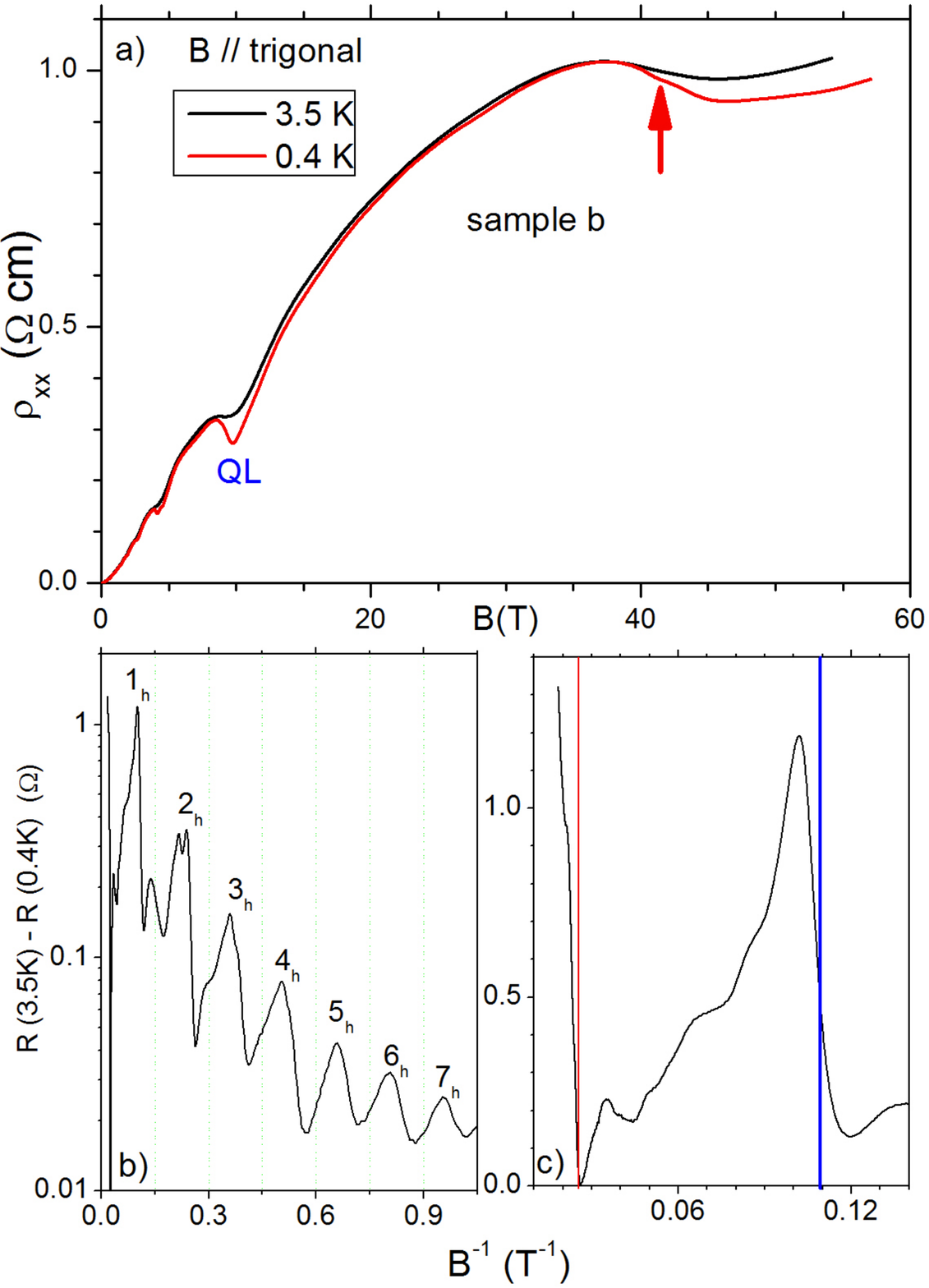}}
\caption{\label{Fig2} a) Resistivity as a function of magnetic field for two different temperatures in another Bi single crystal. The red arrow points to the anomaly resolved here for the first time. b) The field dependence of the resistance differential between 0.4 K and 3.5 K in a semi-logarithmic plot. Quantum oscillations are indexed. c) A linear zoom on the high-field regime. The vertical lines are the same as in Fig. 1c.}
\end{figure}

Fig. 1 presents the Nernst data. The transverse voltage generated by a constant thermal gradient at $T=1.2 K$ is plotted as a function of magnetic field. Larger noise at higher fields is due to the mechanical vibration produced by water cooling of the resistive component of the hybrid magnet.
As previously reported in a study of another single crystal\cite{behnia1}, the data beyond the quantum limit is not featureless and several Nernst peaks are visible between 9 and 35 T. The field positions of these peaks are close but not identical to what was found in the earlier study\cite{behnia1}.  The main new finding is the detection of a new Nernst anomaly centered at B = 37 T.

Nernst effect in bismuth is a probe of quantum oscillations with unmatched sensitivity\cite{behnia2}. As seen in Fig. 1b, plotting the data as a function of the inverse of the field, quantum oscillations with a period of 0.15 T$^{-1}$ are clearly visible. This period corresponds to the extremal cross section of the hole ellipsoid. Below 9T, most of the Nernst peaks can be identified as the intersection between the n-th hole Landau level ($n_{h}$) and the chemical potential. With the approach of the quantum limit (that is, the $1_{h}$ anomaly) additional Nernst peaks appear between the identified ones. As can be seen in the figure, the large anomaly at  37 T anomaly, the focus of this paper, leads to a fourfold increase in the Nernst response. This is almost as drastic as the change caused by the crossing of the hole quantum limit at 9 T and much larger than other unidentified Nernst anomalies.

Resistivity measurements in pulsed magnetic are presented in Fig. 2. Longitudinal resistivity, $\rho_{xx}$, does not show any obvious feature between 9 T and 33 T\cite{behnia1}, but it begins to decrease when the field exceeds 37 T and presents a minimum at a field of about 45 T. This 40 T anomaly becomes more pronounced as the sample is cooled down in a manner comparable to the effect of temperature on the 9 T anomaly.

In bismuth, Shubnikov- de Haas (SdH) oscillations of resistivity superpose on a huge monotonous background\cite{bompadre,behnia2}.  Plotting the difference between the resistance at two temperatures, $ \delta R = R(T=3.5 K)- R(T=0.4 K)$, allows to eliminate this background\cite{yang2}. As seen in Fig. 2b, the SdH oscillations become clearly visible in the dependence of $ \delta R$ on the inverse of the magnetic field. The period of oscillation of  $ \delta R$ (0.15 T$^{-1}$) is identical to the one seen in the Nernst data. The high-field zoom in Fig. 2c shows that the field dependence of  $ \delta R$ may present weak anomalies in the field range between 9 T and 40 T. However, only the latter two field scales are associated with sizeable changes in charge conductivity of comparable magnitude. It is also noteworthy that in the whole field range above 1 T, $ \delta R$ is positive. In other words, the system is weakly metallic. Each SdH oscillation is associated with a periodic enhancement of this metallicity. When B=37 T, $\delta R$ becomes unmeasurably small and then starts another increase.

\begin{figure}
\resizebox{!}{0.34\textwidth}{\includegraphics{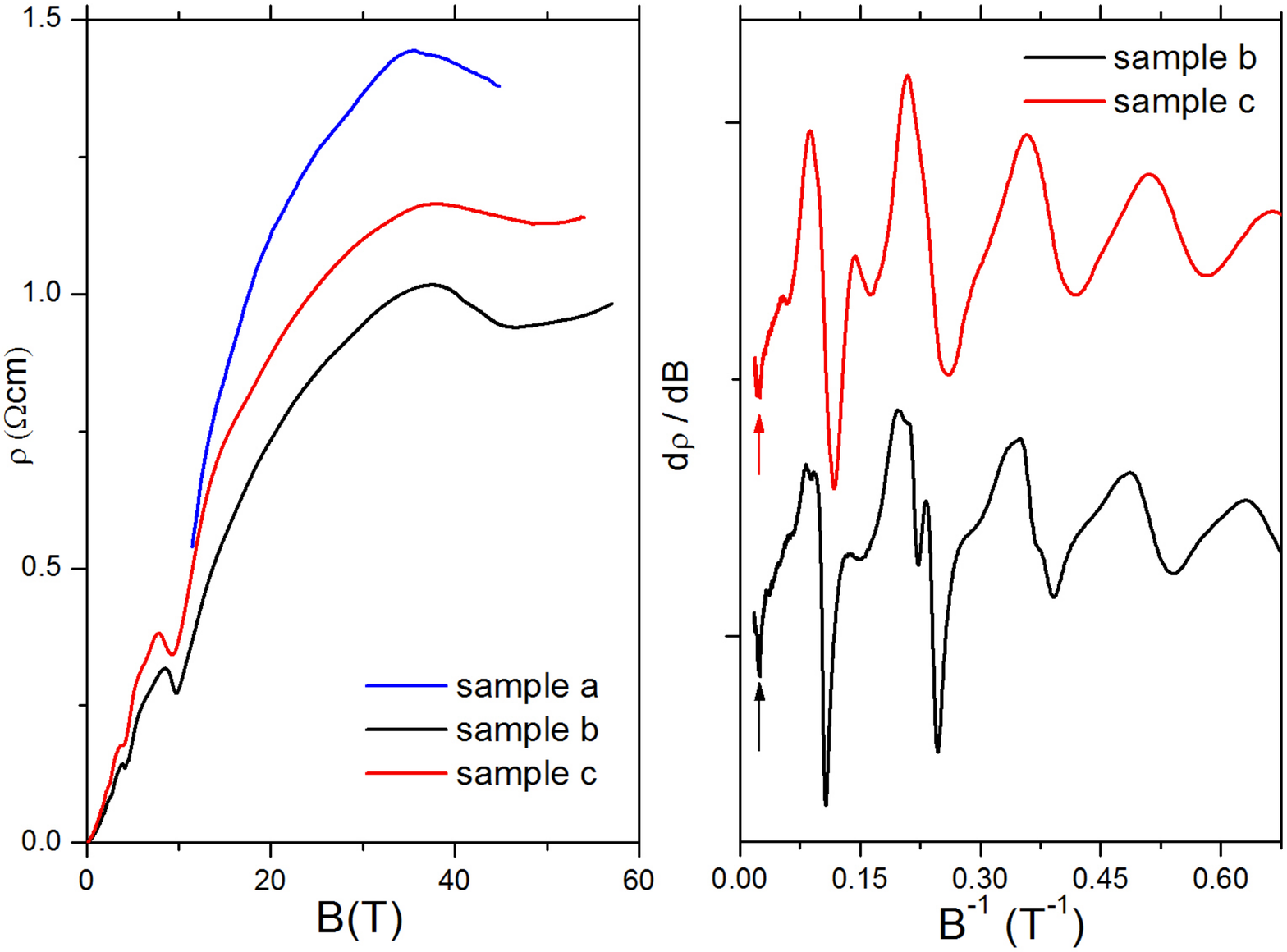}}
\caption{\label{Fig3} a) Magnetoresistance in three different bismuth crystals. Measurements on sample a were performed at 1.2 K in the DC hybrid magnet up to 45 T in parallel to the study of the Nernst effect. Sample b (T=0.4 K) and sample c (T=1.5 K) were measured in a pulsed magnet up to 55 T. b) The field-derivative of the magneto-resistance revealing quantum oscillations. Arrows present the position of the high-field anomaly.}
\end{figure}

This 40 T anomaly was apparently missed in a previous study of high-field magnetoresistance in bismuth, which was carried out at 4.2K and up to 41 T\cite{hiruma}. We checked its presence in five single crystals from different sources. In all of them, the resistivity was found to decrease at a field of about 40 T and this drop became sharper with decreasing temperature. Fig. 3 presents the data in three different crystals. In marked contrast with the data of ref. \cite{hiruma}, in our data the magnetoresistance becomes sublinear above 30 T. The crystals studied here had a resistivity ratio (RRR) in the range of 47 (sample b) to 110 (sample a). No correlation between sample quality and the magnitude or the profile of magnetoresistance was detected.

The existence of such a field scale raises two immediate questions: Can it be the signature of any energy scale associated with the one-particle spectrum? Or can it be the result of a field-induced change in the ground state?

In order to address the first question, we performed angular-dependent resistivity measurements on a bismuth single crystal. Fig.4 presents the results obtained on a third single crystal (sample c). Both longitudinal and Hall resistivity were measured as the magnetic field was tilted off the  trigonal axis in the (trigonal, bisectrix) plane. As seen in panels a and b of Fig.4, the fine structure evolves as the field rotates. In particular, as reported previously\cite{fauque}, there are specific orientations along which, $\rho_{xy}$ is a flat function of magnetic field in the neighborhood of 20 T. However, the 40 T drop in both $\rho_{xx}$  and $\rho_{xy}$ does not disappear, and the change in its field position, if any, is modest.

In the context of the first question, this observation is extremely important. According to the one-particle theory, there are two field scales associated with the electron pockets (anisotropic ellipsoids slightly tilted off the equatorial plane), which are extremely sensitive to field orientation in the vicinity of the trigonal axis\cite{alicea,sharlai}. The torque anomalies observed experimentally\cite{li} track a field scale which tightly follows the theoretical line for $0^{+}_{e}$ Landau levels\cite{sharlai}. More recently, the angular-dependent resistivity studies\cite{fauque} have detected minima in $\rho_{xx}(\theta)$ tracking similar quasi-vertical line in the (B,$\theta$) plane. Clearly, this is not the case of the 40 T anomaly resolved here, which does not show any sizeable dependence on the precise orientation of the magnetic field off the trigonal axis. From this observation follows that this  field scale could not be associated with the single-particle spectrum of the very anisotropic electron pockets. As for the hole pocket, all carriers are confined to the lowest possible Landau level ($0^{-}_{h}$) and due to the large Zeeman energy, this Landau level is not expected to intersect the chemical potential at any magnetic field.

Fig. 4c presents the angular dependence of $\rho_{xx}$ for different magnetic fields. Two minima off the trigonal axis are observable as previously reported in a previous study with a DC field up to 35 T\cite{fauque}. As recalled above, they track the field scale associated with the electron pockets. Interestingly, these two minima are still present in  $\rho_{xx}(\theta)$ obtained for fields exceeding $40 T$. This means that the quasi-horizontal field scale resolved here and the quasi-vertical field scales detected previously\cite{li,fauque} intersect each other as seen in Fig. 4d.

\begin{figure}
\resizebox{!}{0.36\textwidth}{\includegraphics{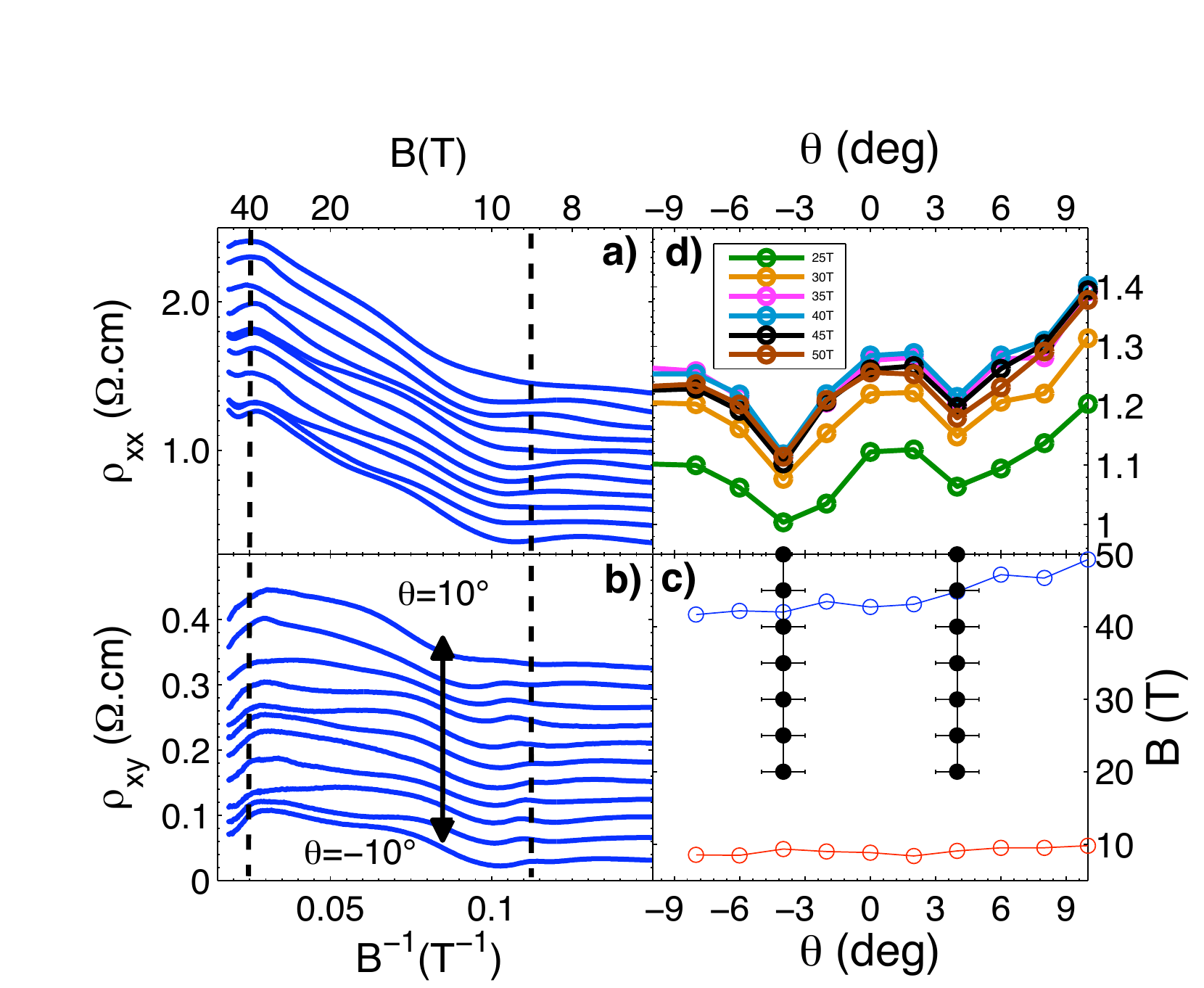}}
\caption{\label{Fig4} Longitudinal (a) and Hall (b) resistivity  at T=1.5 Kas a function of inverse of magnetic  magnetic field in  sample c for different tilt angles in the (trigonal, bisectrix) plane . The two dotted vertical lines at 9 T and 40 T mark the position of the quantum limit and the high-field anomaly when $\theta$=0. Curves are shifted for clarity. c) Angular dependence of the longitudinal resistivity for different magnetic fields at T=1.5 K. Note the presence of two minima. d) The phase diagram, showing the position of three set of anomalies in the (field, $\theta$) plane. The minima in the angular magnetoresistance (solid black circles) track two quasi-vertical field scales off the trigonal axis. The minima in d$\rho$/dB (open blue and red circles) trace two quasi-horizontal lines. The lower line is the quantum limit of the hole pocket. The upper line represents the new field scale found by this study.}
\end{figure}

Li $\emph{et al.}$ \cite{li} have suggested that the quasi-vertical lines may mark distinct phase transitions of electron pockets. The hysteretic jump of magnetization\cite{li} supports such an interpretation. However, the very existence of such a field scale is expected in the one-particle picture\cite{alicea,sharlai}. As discussed above, this is not the case of the 40 T  anomaly. This field scale is not expected in the band picture and there is a fundamental argument against its attribution to the lowest hole ($0^{-}_{h}$) Landau level.  No matter how large the field-induced change in the lattice parameters or the Zeeman energy, if this level intersects the Fermi level alone, the compensated semi-metal would lose its charge neutrality.

It is natural, therefore, to consider this field scale a result of collective interactions between electrons. Theoretically, a three-dimensional electron gas pushed beyond the quantum limit becomes quasi-one-dimensional\cite{yakovenko}. A variety of instabilities have been imagined in such a context\cite{halperin,tesanovic}. Most of the ground states proposed (spin, charge or valley density-wave states, an excitonic insulator or a Wigner crystal), are states commonly associated with a lower conductivity and a worse metallicity, since this would be the usual (but not unavoidable) consequence of the opening of an energy gap.

On the experimental side, our observation is to be compared with the case of the high-field transition observed in graphite\cite{tanuma,iye1}. The jump in magnetoresistance at $B^{*}\simeq30 T$ is commonly believed to be associated with a Charge Density Wave(CDW) transition caused by many-body effects\cite{yoshioka}.  Non-linear charge conductivity measurements supports the CDW picture\cite{iye2}. However, in contrast to what was theoretically expected, the CDW wave-vector was found to be perpendicular to the orientation of the magnetic field. The temperature dependence of the threshold field is in fair agreement with the theoretical prediction\cite{yoshioka}. The effect of the 40 T electronic instability on in-plane electric transport is visibly different in bismuth, where resistivity decreases instead of increasing. Future transport measurements with a current applied along the magnetic field would be instructive. We note that bismuth and graphite differ in several ways. The topology of the Fermi surface is more complex in bismuth and in absence of magnetic field, it is almost isotropic in contrast to the quasi-two-dimensional graphite.

Intriguingly, the signatures of high-field instability in bismuth in both resistivity and Nernst response are comparable to the passage of one-particle Landau levels.  In this respect, this puzzle is linked to the one raised by the observation of the smaller Nernst anomalies in both Bi\cite{behnia1} and Bi$_{0.96}$Sb$_{0.04}$\cite{banerjee} beyond the quantum limit. The explanation of an ultraquantum field scale lie out of the scope of the band picture and implies an appropriate treatment of the electronic interactions, which are expected to become large as the quantum limit is crossed\cite{halperin,alicea}.

In two dimensions, such interactions lead to the emergence of the Fractional Quantum Hall Effect out of the Integer Quantum Hall Effect. In our three-dimensional context, in contrast to the two-dimensional case, the longitudinal resistivity is finite and the Hall resistivity, which presents plateaus for particular orientations of the magnetic field \cite{fauque}, is not quantized. These features indicate a different departure point. The challenge to many-body quantum theory of electrons is to determine how fundamental this difference is and to conceive collective electronic effects in three-dimensions giving rise to field scales other than those associated with integer Landau levels. Very recently, such theoretical explorations has been carried out\cite{burnell,levin}.

We thank L. Balicas and T. Murphy for their technical assistance and J. Alicea, L. Balents, T. Giamarchi, Y. Sharlai and V. Yakovenko for useful discussions. This work is supported by the Agence Nationale de Recherche as (DELICE: ANR-08-BLAN-0121-02). LNCMI-Toulouse is supported by EuroMagNET. Research at the NHMFL is supported by the NSF, by the State of Florida and by the DOE.

\end{document}